\documentclass[11pt,onecolumn,amssymb,nofootinbib]{revtex4}
\usepackage{amsmath, amsthm, amscd, amssymb}
\usepackage{bm}
\usepackage{bbm}

\begin{document}
\title{\bf Connecting the dots:  Mott for emulsions, collapse models, colored noise, frame dependence of measurements, evasion of the ``Free Will Theorem'' }

\author{Stephen L. Adler}
\email{adler@ias.edu} \affiliation{Institute for Advanced Study,
Einstein Drive, Princeton, NJ 08540, USA.}

\begin{abstract}
We review the argument that latent image formation is a measurement in which the state vector collapses,
requiring an enhanced noise parameter in objective reduction models.  Tentative observation of a residual
noise at this level, plus several experimental bounds, imply that the noise must be colored (i.e., non-white), and hence frame dependent and non-relativistic. Thus a relativistic objective reduction model, even if achievable in principle,  would be incompatible with experiment; the best one can do is the
non-relativistic CSL  model. This negative conclusion has a positive aspect, in that the non-relativistic CSL reduction model evades the argument leading to the Conway--Kochen  ``Free Will Theorem''.
\end{abstract}
\maketitle

\section{The Mott scenario applied to an emulsion stack}

In a celebrated paper \cite{mott} dating from the early days of quantum theory,  Neville Mott answered the question
of how a spherically symmetric $\alpha$ decay in a cloud chamber can lead to a linear track. His argument is based on excitation of two or more atoms to form the track, but the core of Mott's calculation can be
stated \cite{adler1} using just one atom.  Consider an alpha particle emitted in a spherical wave by a nucleus located at the coordinate origin, and an atom at location $\vec a$.  We want the amplitude for the
alpha particle to scatter to a plane wave with wave number $\vec k$, conditional on the atom being excited from an initial state $\psi_0$ to an excited state $\psi_S$.
For an alpha particle with coordinate $\vec R$ and an atomic electron with coordinate $\vec r$,
and assuming the atom excitation energy is negligible,  the Born approximation (dropping overall constants) is proportional to
\begin{equation}\label{born1}
f(\vec k) \propto \int d^3R d^3r ~\psi_{\rm final}^*  V \psi_{\rm initial} ~~~,
\end{equation}
with
\begin{align}
\psi_{\rm final}^* \propto& e^{-i\vec k\cdot \vec R} \psi_S^*(\vec r)~~~,\cr
V\propto &1/|\vec R-\vec r|~~~,\cr
\psi_{\rm initial} \propto& \big(e^{ik|\vec R|}/|\vec R|\big) \psi_0(\vec r)~~~.\cr
\end{align}
We thus  get
\begin{equation}\label{born2}
f(\vec k) \propto \int d^3 R R^{-1} e^{ik R(1-\hat k \cdot \hat R)} V_{0S}(\vec R)~~~,
\end{equation}
where $\vec k =k \hat k$,  $\vec R = R \hat R$, and where following Mott we have defined
\begin{equation}
V_{0S}(\vec R)= \int d^3 r \psi_S^*(\vec r) \psi_0(\vec r)/|\vec R-\vec r|~~~.
\end{equation}
Since $V_{0S}(\vec R)$ is significantly different from 0 only for $\vec R \simeq \vec a$,
we can approximate Eq. \eqref{born2} by making the replacement
 $\hat R \simeq \hat a$, where $\vec a =|\vec a|\hat a$, giving
\begin{equation}\label{born3}
f(\vec k) \propto \int d^3 R R^{-1}  e^{ik R(1-\hat k \cdot \hat a)} V_{0S}(\vec R)~~~.
\end{equation}
Mott then notes that the coefficient of $R$ in the exponent is rapidly oscillating except when
$1 \simeq \hat k \cdot \hat a$, that is when the outgoing wave vector is nearly parallel to the vector from
the alpha emitter to the first atom.  Thus a second atom can be excited only if it lies in
a small cone about a vector $\hat a$ pointing from the first atom, since the fact that the first
atom was excited means that the subsequent scattered wave has its origin at this atom.  Mott's
more detailed treatment calculates the form of the wave scattered by the first atom, but the
magnitude is governed by the amplitude of Eq. \eqref{born3} above.

Although Mott's argument was motivated by tracks in cloud chambers, it applies equally well to an $S$-wave
alpha emitter embedded in an emulsion stack.  The difference is that, unlike cloud chamber
detection, where drop nucleation and drop expansion are temporally inseparable,  in an emulsion latent image formation can be separated by as much time as desired from the subsequent development which renders the track visible.  One can then ask when does state vector collapse occur, in latent image formation or in the subsequent development?  Since the initial state is spherically symmetric, for collapse to occur only
on development would require that a spherically symmetric  infinity of virtual tracks be present up to
the exposure of the emulsion plate to a developer.  This makes no sense both in terms of the molecular  mechanism  of latent image formation, and in terms of energetics.\footnote{In a stimulating email correspondence, Jerry Finkelstein has remarked that energetics would allow a spherical coherent superposition of latent image tracks, each with the energetics of formation of a single track, with the actual track that is formed picked out later by development.  But this is not compatible with the standard, and verified, picture of latent image formation as a physical process in which atoms have moved, quite independent of the larger atomic motions that take place much later on development.} So it
seems inescapable that when the Mott scenario is applied to an alpha emitter embedded in an emulsion
stack, state vector reduction already occurs at latent image formation. This reinforces the assertion, made  long ago  by Gisin and Percival \cite{gisin}, that
``formation of the latent image in photography...shows unequivocally that amplification up to the macroscopic level
is quite unnecessary for the formation of a permanent classical record of a quantum event...''.

\section{Noise coupling strength in objective reduction models needed to make latent image formation a mesurement}

This conclusion has significant consequences for so-called ``collapse models'', in which state vector
collapse in measurement is the result of a definite physical process that modifies linear Schr\"odinger
dynamics. In collapse models \cite{collapse}, and in particular the extensively studied continuous spontaneous localization (CSL)
model, a small noise term is nonlinearly coupled to the Schr\"odinger equation, in a way that is made
unique by the physically plausible requirements of preservation of state vector normalization, and absence
of faster-than-light signaling.  In  processes in which few atoms move, the noise has little effect and
linear Schr\"odinger dynamics obtains.  But in processes in which a microscopic system is entangled with
a massive apparatus, many atoms are involved and the noise effects accumulate.  The initial estimates
for CSL noise coupling assumed as a typical apparatus a small pointer, and derived a suggested noise
coupling strength based on requiring state vector reduction when the pointer moves by a macroscopically
observable amount.  However, in latent image formation far fewer atoms move than in the pointer model,
and a correspondingly larger noise strength is needed.  This was calculated by Adler \cite{adler2} based on
a survey of detailed models for photographic image formation, with
the result that for state vector reduction to occur in latent image formation, the noise coupling must
be a factor of about  $2 \times 10^{9\pm 2}$ times larger than in the pointer model of measurement, for
the standard assumption of a noise correlation length $r_C=10^{-5}$ cm.   There are also
upper bounds on the noise strength coming from observed limits on the anomalous heating produced by
CSL noise.  The best of these bounds comes \cite{adler2} from an analysis of heating  of the inter-galactic
medium (IGM) , leading to the conclusion that for latent image formation to constitute a measurement, and for consistency with the observed IGM heating over the age of the universe, the CSL
noise must be a factor of $10^{8\pm 1}$ times larger than the estimates from the pointer model of
measurement, corresponding to a CSL noise coupling strength $\lambda$ of
\begin{equation}\label{noise}
\lambda \sim 2 \times 10^{-9\pm 1}
{\rm s}^{-1}~~~.
\end{equation}
Small complex-number-valued fluctuations in the spacetime metric $g_{\mu\nu}$ have been suggested \cite{adler3}, \cite{gasb} as a plausible physical origin
for the noise.

\section{The noise, if present, must be colored, and thus frame dependent}

Let us now assume for further discussion that a CSL noise with coupling strength of order Eq. \eqref{noise} is present.  This is consistent with the tentative finding by Vinante et al. \cite{vin} in a cantilever experiment of a residual noise with
coupling strength $\lambda=10^{-7.7} {\rm s}^{-1}$, and with the CSL noise bound coming from the LISA pathfinder mission \cite{carlesso}, \cite{helou}.    A number of empirical constraints imply that
this noise cannot be white noise, that is, it must be ``colored'' noise with a spectral  frequency cutoff.  The first is the non-observation of noise-induced gamma ray emission from germanium \cite{germ}, which implies that the noise strength of order Eq. \eqref{noise}  is possible only if
the noise power spectrum cuts off at an angular frequency below $\sim 15 \,{\rm keV}/\hbar \sim 2 \times 10^{19} {\rm s}^{-1}$, reflecting the lower end of the gamma ray frequency range that is searched.  The others are bounds on bulk heating reviewed by Adler and Vinante \cite{avn}, coming from limits on unexplained bulk
heating in both cryogenic experiments and in the Earth energy balance.   These show that the effective noise coupling $\lambda_{\rm eff} $ for bulk heating must be less than around $ \sim 10^{-11} {\rm s}^{-1}$.   Since $\lambda_{\rm eff}$ for heating through phonon emission is related to the noise power spectrum  $\lambda(\omega)$ by \cite{avn}, \cite{bahrami}
\begin{equation}\label{lameff}
\lambda_{\rm eff}= \frac{2}{3 \pi^{3/2}} \int d^3w e^{-\vec w^2} \vec w^2 \lambda(\omega_L(\vec w/r_c))~~~,
\end{equation}
where $\omega_L(\vec q)$ is the longitudinal phonon frequency at wave number $\vec q$, the bulk heating  bound on $\lambda_{\rm eff}$ translates into a noise frequency cutoff of around
 $\omega_L(|\vec q|=r_c^{-1} )\sim v_s  r_c^{-1} \sim 0.4 \times 10^{11} {\rm s}^{-1}$, with  $v_s$ the speed of sound, which for this estimate is  taken as $4\, {\rm km}/{\rm s}$.  Thus, if the noise reported in \cite{vin} at the
very low cantilever frequency of $8174 ~{\rm s}^{-1}$ is an observation of CSL noise, it provides further
evidence that the noise power spectrum must have a high frequency cutoff, and hence be non-white or
``colored'' noise.

This has important implications for the relativistic invariance of the theory.  White noise is characterized by a random variable $\frac{dW_t(\vec x)}{dt}$ with expectation
\begin{equation}\label{noiseex}
{\cal E}\left[\frac{dW_t(\vec x)}{dt}\, \frac{dW_{t^{\prime}}(\vec y)} {dt^{\prime}}\right]
 \propto \delta(t-t^\prime) \delta^3(\vec x-\vec y)~~~,
\end{equation}
which is Lorentz invariant,  making  it possible to envisage a  relativistically invariant collapse model, as recently discussed by  Pearle \cite{pearle} and  by Bedingham
\cite{bedingham}.  However, for non-white noise the analogous noise expectation is
\begin{equation}\label{noiseex1}
{\cal E}\left[\frac{dW_t(\vec x)}{dt}\, \frac{dW_{t^{\prime}}(\vec y)} {dt^{\prime}}\right] \propto \frac{1}{2\pi} \int_{-\infty}^{\infty} d\omega
\lambda(\omega) e^{-i\omega(t-t^{\prime})}\delta^3(\vec x-\vec y)~~~.
\end{equation}
If $\lambda(\omega)$ is not a constant, but rather has a high frequency cutoff,
such as $\lambda(\omega)=e^{-\omega^2 t_C^2}$ with $t_C$ a correlation time,
the noise expectation of Eq. \eqref{noiseex1} is no longer Lorentz invariant. In the CSL model it is
customary to include a spatial smearing in coupling the noise to a system operator such as the mass density. If this smearing is interpreted as a noise correlation length,  its inclusion in Eq. \eqref{noiseex1} gives a  two-parameter form for the noise correlation, with both a spatial correlation
length $r_C$ and a correlation time $t_C$,
\begin{equation}\label{noiseex2}
{\cal E}\left[\frac{dW_t(\vec x)}{dt}\, \frac{dW_{t^{\prime}}(\vec y)} {dt^{\prime}}\right] \propto \frac{1}{(2\pi)^4} \int_{-\infty}^{\infty} d\omega \int d^3 q e^{-\omega^2 t_C^2-\vec q^2 r_C^2} e^{i[\vec q \cdot (\vec x-\vec y)- \omega(t-t^{\prime})]}~~~,
\end{equation}
which again is manifestly not Lorentz invariant.

Thus, if experiment shows that  CSL is the correct resolution of the quantum measurement problem, by giving a mechanism
for objective reduction of the state vector, there cannot be a relativistic generalization.  The theory is intrinsically non-relativistic, and picks a preferred Lorentz frame. In writing Eq. \eqref{noiseex2} we have implicitly assumed that the noise is spatially isotropic, that is, both translationally and rotationally invariant.  A boost with velocity $\vec v$ from the frame in which the noise is isotropic will modify Eq. \eqref{noiseex2} to have a dependence on $\vec v$.  Although the idea of a preferred frame
seems counterintuitive to one steeped in special relativity theory, there is already experimental indication of a preferred frame in Nature, the rest frame of the
cosmic microwave blackbody (CMB) radiation.  The simplest assumption about the CSL noise is that its preferred rest frame, in which it is isotropic, coincides with the
CMB rest frame.  Results of a measurement in any other frame can then be inferred by making a Lorentz boost from what is observed in this preferred frame.  For example, in a frame boosted by velocity $\vec v$ from the CMB frame, the exponent $i[\vec q \cdot (\vec x-\vec y)- \omega(t-t^{\prime})]$ in Eq. \eqref{noiseex2} becomes
\begin{equation}\label{noiseex2shifted}
i\vec q_\perp \cdot (\vec x_\perp-\vec y_\perp) + i \gamma [(q+\omega v /c^2)(x-y)-(\omega + qv)(t-t^\prime)] ~~~,
\end{equation}
with $q,\,x,\,y$ denoting components parallel to the direction of the boost, and $\gamma=(1-v^2/c^2)^{-1/2}$. This leads to a direction dependence of the spatial and temporal correlations that has not  been systematically incorporated into phenomenological studies (apart from investigations \cite{toros} of the
 dissipative CSL model), and may be significant because the  velocity of the solar system relative to the CMB is nearly 400 km/s.   Alternatively, if the spatial smearing is treated as a Lorentz scalar and not interpreted as a correlation length in the noise, the effect of a boost on Eq. \eqref{noiseex1} is to replace $-i\omega (t-t^\prime)$  by $-i\gamma\omega (t-t^\prime)$. This is the
frame-dependent effect noted by Shan Gao in the context of the energy conserving reduction
model developed in his book \cite{shan1}, where he has made the proposal that there is a non-white noise driving reduction, which picks the CMB rest frame as a preferred frame, and which leads to detectible effects in other Lorentz frames.

Given that the CSL noise, in the form either of Eq. \eqref{noiseex1} or Eq. \eqref{noiseex2}, is frame
dependent, does the CSL model give a theory of measurement that is  robust under boosts with
$v/c<<1$? The answer is both ``yes'' and ``no''.  ``Yes'' because, since the proofs of
state vector reduction in CSL apply for general classes of noises,  the theory in any low velocity frame is expected to yield the correct statistical properties of measurement as dictated by the Born and
L\"uders rules,  which is the minimal requirement for an acceptable theory. ``No''  because, since the noise determines which component of a superposition is
picked as the outcome of any given individual
run of an experiment, the fact that the noise changes with change of Lorentz frame suggests that individual
outcomes can differ when the experiment is performed in different frames starting from the same initial conditions\footnote{We emphasize, by ``performed in different frames'' we specifically mean  with the apparatus boosted to different frames.  Boosting the
apparatus to a new frame is different
from performing an experiment in a fixed frame and viewing it from boosted frames, which gives
the same result regardless of whether the physics of the experiment is Lorentz invariant or not, as long
as ``viewing'' refers to use of electromagnetic radiation, which obeys the Lorentz covariant Maxwell
equations.}, just as outcomes can differ in repetitions
of the experiment in the same frame.   Only the statistical aspects of repeated measurements are expected to be robust under changes of frame, not the sequence of
individual outcomes.

\section{The nonrelativistic CSL model evades the ``Free Will Theorem''}

In a striking series of papers based on assuming relativistic invariance of measurements, Conway and Kochen  \cite{conway1}, \cite{conway2}, \cite{kochen} proved their  ``Free Will Theorem''  (FWT), which implies  that a  relativistic  objective theory of measurement cannot exist: ``Granted our three axioms, the FWT shows that nature itself is non-deterministic.  It follows that there can be no correct relativistic
deterministic theory of nature....Moreover, the FWT has the stronger implication that there can be no relativistic theory that provides a mechanism for reduction.''  We shall now argue that this conclusion
does not generalize to the nonrelativistic CSL model, both because this model is nonlocal, and because the
noise that drives state vector reduction is frame dependent.

To begin, let us recall the axioms on which the FWT is based.  Two of them, SPIN and TWIN,
are standard quantum mechanical statements about the properties of spin 1 measurements (SPIN) and the properties of pairs of identical particles with nonzero spin entangled in a spin 0 state (TWIN).  A third
essential axiom involves the assumption of relativistic invariance, and is given in  different forms
in the three FWT theorem papers cited above.  In \cite{conway1}, relativity is introduced through an axiom
FIN, ``There is a finite upper bound to the speed with which information can be effectively transmitted.  This is, of course, a well-known consequence of relativity theory, the bound being the speed of light.''
The authors go on to say that ``FIN is not experimentally verifiable directly .....Its real justification  is that it follows from relativity and what we call ``effective causality,'' that effects cannot proceed
their causes.''  In \cite{conway2} FIN is replaced by an axiom MIN, which also  embodies causality.
The authors note that ``One of the paradoxes introduced by relativity was the fact that temporal order
depends on the choice of inertial frame.  If two events are spacelike separated, then they will appear in one time order with respect to some inertial frames, but in the reverse order with respect to others.... It is usual tacitly to assume the temporal causality principle that the future cannot alter the past.  Its
relativistic form is that an event cannot be influenced by what happens later in any given inertial frame.''  The MIN axiom, based on these statements, but referring specifically to elements of the proof of their
theorem, states: ``Assume that the experiments performed by A and B are space-like separated.  Then
experimenter B can freely choose any one of the 33 particular directions w, and a's response is independent
of this choice.  Similarly and independently, A can freely choose any one of the 40 triples x, y, z and b's
response is independent of that choice.''  Finally, the most recent paper, Kochen \cite{kochen}, postulates relativistic
invariance through an axiom LIN,  which  ``states that the result of an experiment is Lorentz covariant: a
change of Lorentz frames does not change the results of the experiment''.  Even though the earlier papers
\cite{conway1} and \cite{conway2} introduce relativity through the alternative axioms FIN and MIN, the idea
of frame independence of measurement outcomes is implicit in the proofs given in these papers as well.

\subsection{The nonrelativistic CSL model does not obey FIN and MIN}

To see that the nonrelativistic CSL model does not obey FIN and MIN, it is not necessary to take account
of the non-white or colored nature of the noise.  Since the details of CSL with  non-white noise \cite{adler-bassi} are more complicated than those of CSL with white noise, we shall focus in this
section on the CSL model with white noise, which can be analyzed using the It\^o calculus for describing
the noise.  The white noise CSL model is defined by the following equations.  The state vector $|\psi(t)\rangle$ evolves in time according to the nonlinear stochastic differential equation
\begin{equation}\label{CSLeq}
d|\psi(t)\rangle=\big[-\frac{i}{\hbar} H dt + \gamma_{CSL}^{1/2}\int d^3x \big(M(\vec x)-\langle M(\vec x) \rangle\big) dB(\vec x)
-\frac{\gamma_{CSL}}{2} \int d^3 x  \big(M(\vec x)-\langle M(\vec x) \rangle\big)^2 dt \big]|\psi(t)\rangle~~~.
\end{equation}
Here $H$ is the Hamiltonian, $ dB(\vec x)$ is a Brownian motion obeying
\begin{equation}\label{brownian}
dt dB(\vec x)=0~~,~~~dB(\vec x)dB(\vec y)=\delta^3(\vec x-\vec y) dt~~~,
\end{equation}
$\langle M(\vec x)\rangle$ is the expectation of $M(\vec x)$ in the state  $|\psi(t)\rangle$,
$M(\vec x)$ is the mass density operator smeared over a normalized spatial Gaussian $e^{- \vec x^2/(2 r_C^2)}$, and $\gamma_{CSL} =8 \pi^{3/2} r_C^3 \lambda$ is the noise coupling strength.   One can readily check
that because of the nonlinear term in Eq. \eqref{CSLeq} with coefficient $\gamma_{CSL}$, the evolution of Eq.
\eqref{CSLeq} is norm preserving, that is, $d\langle \psi(t)|\psi(t)\rangle =0$.

The paradigm for the TWIN axiom of the FWT is the Einstein-Podolsky-Rosen (EPR)\cite{epr} experiment with spin-$\frac{1}{2}$ particles correlated in a momentum $\vec 0$ spin singlet state, with the particles
moving apart with finite velocities $\pm\vec V$.  Labeling the particles $A$ and $B$, the state vector
before measurement is
\begin{equation}\label{before}
|\Psi\rangle = 2^{-1/2}[ |A \uparrow\rangle |B \downarrow\rangle-|B \uparrow\rangle |A \downarrow\rangle]~~~.
\end{equation}
After the particles are well separated, a Stern-Gerlach apparatus to measure spin along an axis $\vec x$ is
inserted in the path of particle $A$, leading to the entangled state vector
\begin{equation}\label{before1}
|\Psi \rangle = 2^{-1/2}[ |A \uparrow\rangle |B \downarrow\rangle|{\rm apparatus} \uparrow\rangle-|B \uparrow\rangle |A \downarrow\rangle|{\rm apparatus} \downarrow\rangle]~~~.
\end{equation}
Because the apparatus is massive, the Brownian noise leads to rapid reduction of the state vector to either
\begin{equation}\label{after1}
|A \uparrow\rangle |B \downarrow\rangle|{\rm apparatus} \uparrow\rangle
\end{equation}
 or
\begin{equation}\label{after2}
|B \uparrow\rangle |A \downarrow\rangle|{\rm apparatus} \downarrow\rangle~~~;
\end{equation}
note that these are normalized to unity because the CSL evolution is norm preserving.  Immediately
after this reduction, because of the non-locality implicit in the CSL equation, the state vector  of $B$ is fixed to be either spin down or spin up.    This should not be thought of as a noise
signal running from the measurement at $A$ to $B$; in fact, because of the locality of the noise in
Eq. \eqref{brownian} and the short range of the correlation length $r_C$, the action of the noise at
$B$ is essentially independent of that at $A$, and serves just the function of preserving the state vector norm.
So in the CSL model, since the wave function is regarded as an element of physical reality, there
is an effectively instantaneous change in the state of $B$ once the measurement is made at $A$.  From
this point on, any spin measurement performed by $B$, along any axis, is governed by $B$'s reduced
state vector.

To see that this violates both FIN and MIN,  let us recall what relativity says about the temporal ordering
of events $\vec x_A$ at time $t_A$ and $\vec x_B$ at time $t_B$ in a ``rest'' frame, when viewed from a boosted frame moving with velocity $v<c$ along the axis $\vec x_A-\vec x_B$.  Since we are dealing with only one
axis, we can drop the vector arrows and denote the events as $x_A,\, t_A$ and $x_B,\, t_B$.  From the
Lorentz transformation to the boosted frame,
\begin{equation}\label{Ltran}
x^\prime=\gamma(x-vt)~~,~~~t^\prime=\gamma(t-vx/c^2)~~,
\end{equation}
we find for the time interval between the events in the boosted frame
\begin{equation}\label{tdiff1}
t^\prime_B-t^\prime_A=(t_B-t_A) \gamma (1-vv_{AB}/c^2)~~~,
\end{equation}
where we have defined an effective velocity relating the two events as
\begin{equation}\label{veff}
v_{AB}=(x_B-x_A)/(t_B-t_A)~~~.
\end{equation}
We see that if $|v_{AB}|<c$, then the temporal ordering of the events is the same in the rest frame
and all boosted frames.  But if $|v_{AB}|>c$, as in the CSL treatment of the EPR experiment, not only is
FIN violated, but we can find boosted frames in which the temporal order of the events is inverted,
giving apparent backwards causation, violating MIN.

\subsection{The nonrelativistic CSL model does not obey LIN}

As we have stressed in Sec. III, in the non-relativistic CSL model
the noise that determines the outcome of any individual measurement changes with Lorentz frame, in contradiction to  the axiom LIN.   To elaborate on this, let us take the frame in which the noise is spatially isotropic as the CMB rest frame. Consider
a Stern-Gerlach experiment in this frame, which measures the $x$ component of a spin that is initially
in a $z$ axis spin up state.  In the CSL model, if the noise coupling is set to zero, the initial superposition of spin up and spin down along the $x$ axis is maintained, and the apparatus entangled with the spin goes into a  ``Schr\"odinger cat''
state in which there is a macroscopic coherent superposition.  With the noise coupling nonzero, the
``cat'' state is reduced to one of its two components, either spin up or spin down along the $x$ axis, as
a result of the action of the noise on the entangled spin--apparatus system.  But since {\it the noise
history picks the outcome}, if the apparatus is boosted to a different frame,
where the noise history is different,  there is no guarantee that the outcome will be the same as when the experiment is at rest in the CMB frame. Note that the noise cannot be regarded as a small perturbation
from the viewpoint of its role in picking a definite experimental outcome; if the noise is made weaker, the
objective reduction produced by the noise takes longer, but the outcome still depends on details of the
noise history, which are not Lorentz invariant.   This violates LIN, which asserts that individual outcomes starting from the same initial conditions should be the same when the
experiment is done in any Lorentz frame boosted with respect to the original frame, and prevents the FWT contradiction of forming a forbidden 101 function.    The proof of the FWT requires that
the outcome of an experiment be independent of the Lorentz
frame in which it is performed, at the level of individual outcomes, not of accumulated statistics:   B  makes only one
measurement for the direction w that B  has chosen, not a sequence of measurements, and A makes only one measurement for each x, y, z of the triple that A has chosen, not a sequence of measurements. So frame independence of the statistical aspects of repeated measurements is not enough to allow proof of the FWT; frame independence of individual outcomes is required.\footnote{Instantaneous
measurements are an idealization, and we leave for future study questions relating to finite measurement times:  (1) With white noise, CSL predicts a measurement rate governed by the noise coupling and
the initial state variance.  For non-white noise, with correlation time $t_C$, does this rate change so that the measurement time $t_{\rm MEAS}$ is greater than $t_C$?  (2) For successive measurements with white noise, different outcomes can be obtained with an infinitesimal time interval between measurements.  With non-white noise,
what is the minimal interval between successive measurements for independent outcomes to be possible?
Is it of order $t_C$?  (3)  According to Eqs. \eqref{tdiff1} and \eqref{veff}, there is a minimum boost
velocity $v$ needed for a reversal of temporal ordering to be possible, given by
$v_{\rm MIN} > c^2/v_{AB}=c^2 (t_B-t_A)/(x_B-x_A)$.  If we assume that $t_B-t_A$ is at least the measurement time at $A$, does the implied condition $v_{\rm MIN} > c^2 t_{\rm MEAS}/(x_B-x_A)$ suffice to guarantee that when $B$ is in the  boosted frame, the noise acting on $B$  can give an outcome different from what would be obtained if $B$ were in the same frame as $A$?}

\section{Summary}

To summarize, as we have emphasized in earlier sections by ``connecting the dots'',
the Mott scenario for emulsions implies that latent image formation is a measurement, requiring an enhanced CSL noise coupling strength, which in turn from existing
experimental bounds implies that the noise (if present at all) {\it must} be non-white, and thus frame dependent.  Hence only a nonrelativistic objective reduction model can be compatible with experiment,
and this in turn allows evasion of the ``no-go'' theorem proved by Conway and Kochen which assumes   relativistic invariance of measurements. So if the tentative noise signal reported in \cite{vin}
is confirmed, Nature will once again, through unanticipated complexity, have  found a way of evading startling paradoxical conclusions.

\section{Acknowldegements}
I wish to thank Angelo Bassi, Jeremy Bernstein, Jerry Finkelstein, Shan Gao, Si Kochen, and Andrea Vinante  for stimulating  conversations and/or email correspondence about the mysteries of quantum measurement. This work was performed in part at the Aspen Center for Physics, which is supported by the National Science Foundation under Grant No. PHY-1607611.

\end{document}